\documentclass[12pt,reqno]{amsart}

\usepackage{epsfig}

\newcommand{\varphijk}{\varphi_{j_k}}
\newcommand{\pdo}{\psi DO}
\newcommand{\Tr}{\operatorname{Tr}}

\newcommand{\oit}{\operatorname{ it}}

\newcommand{\sa}{\sigma_A}

\newsymbol\upharpoonright 1316

\newcommand{\R}{{\mathbb R}}
\newcommand{\C}{{\mathbb C}}

\newcommand{\Z}{{\mathbb Z}}

\newcommand{\N}{{\mathbb N}}

\newcommand{\E}{{\mathbf E}}
\renewcommand{\H}{{\mathbf H}}
\newcommand{\half}{{\frac{1}{2}}}

\renewcommand{\phi}{\varphi}

\newcommand{\ecal}{\mathcal{E}}
\newcommand{\hcal}{\mathcal{H}}

\newcommand{\qcal}{\mathcal{Q}}

\newcommand{\mcal}{\mathcal{M}}

\newtheorem{theo}{{\sc Theorem}}
\newtheorem{prob}{{\sc Problem}}
\newtheorem{theoconj}{{\sc Theorem/Conjecture}}

\title[Quantum ergodicity and mixing of eigenfunctions]
{Quantum ergodicity and mixing of eigenfunctions}

\author{Steve Zelditch}
\address{Department of Mathematics, Johns Hopkins University, Baltimore,
MD 21218, USA}
 \email{szelditch@jhu.edu}

\thanks{Research partially supported by NSF grant
 \#DMS 0302518.}

\date{January 25, 2005}

\begin{document}

\maketitle

\begin{abstract} This article surveys mathematically rigorous
results on quantum ergodic and mixing systems, with an emphasis on
general results on  asymptotics of eigenfunctions of the Laplacian
on compact Riemannian manifolds.

\end{abstract}

Quantum ergodicity and mixing belong to the field of  Quantum
Chaos, which studies quantizations of `chaotic' classical
Hamiltonian systems. The basic questions are, how does the chaos
of the classical dynamics impact on  the eigenvalues
eigenfunctions of the quantum Hamiltonian $\hat{H}$ and on and
long time dynamics generated by $\hat{H}$?

These problems lie at the foundations of the semi-classical limit,
i.e. the limit as the Planck constant $\hbar \to 0$ or the energy
$E \to \infty$. More generally, one could ask what impact  any
dynamical feature of a classical mechanical system (e.g. complete
integrability, KAM, ergodicity) has on the eigenfunctions and
eigenvalues of the quantization.

 Over the last $30$ years or so, these questions
have been studied rather systematically by both mathematicians and
physicists. There is an extensive literature comparing classical
and quantum dynamics of  model systems,  such as comparing the
geodesic flow and wave group on a  compact (or finite volume)
hyperbolic surface, or comparing
 classical and quantum billiards on the Sinai billiard or the Bunimovich
 stadium, or comparing the  discrete dynamical system generated by a  hyperbolic
 torus automorphism and its quantization by the metaplectic representation. As these models indicate, the basic problems
 and phenomena
 are richly embodied  in simple, low-dimensional examples in much
 the same way that two-dimensional toy statistical mechanical models
 already illustrate complex problems on phase transitions.  The
 principles established for simple models should apply to far more
 complex systems such as atoms and molecules in strong magnetic
 fields.

 The conjectural picture which has emerged from many computer
 experiments and heuristic arguments on these  simple model systems is
 roughly that there exists a length scale in which   quantum chaotic systems
 exhibit universal behavior.  At this length scale, the eigenvalues
 resemble eigenvalues of random matrices of large size
 and the eigenfunctions resemble random waves. A small sample  of the
 original physics articles suggesting this picture is \cite{B,
 BGS, FP, Gu, H, A}.

 This article  reviews some of the rigorous mathematical
 results in quantum chaos, particularly
the rigorous results on eigenfunctions of quantizations of
classically ergodic or mixing systems. They support the
conjectural picture of random waves up to two moments, i.e. on the
level of means and variances.  A few results also exist on higher
moments in very special cases. But from the mathematical point of
view,
  the  conjectural links to random matrices or random waves
 remain very much
 open at this time. A key difficulty is  that the
 length scale on which universal behavior should occur is very far
 below the resolving power of any known mathematical techniques,
 even in the simplest model problems. The main evidence for
 the random matrix and random wave connections comes from numerous computer
 experiments of model cases in the physics literature. We will not
 review numerical results here, but to get a well-rounded view of
 the field it is important to understand the computer experiments
 (see \cite{BSS, Bar, KH} for some examples).

The model quantum systems that have been most intensively studied
in mathematical quantum chaos are Laplacians or Schr\"odinger
operators on compact (or finite volume) Riemannian manifolds, with
or without boundary,  and quantizations of symplectic maps on
compact K\"ahler manifolds. Similar   techniques and results apply
in both settings, so for  the sake of coherence we concentrate on
the Laplacian on a compact Riemannian manifold with `chaotic'
geodesic flow and only briefly allude to the setting of `quantum
maps'. Additionally, two main  kinds of methods are in use: (i)
methods of semi-classical (or microlocal) analysis, which apply to
general Laplacians,  and (ii)  methods of number theory and
automorphic forms, which apply to arithmetic models such as
arithmetic hyperbolic manifolds or quantum cat maps. Arithmetic
models are far more `explicitly solvable' than general chaotic
systems, and the results obtained for them are far sharper than
the results of semi-classical analysis. This article is primarily
devoted to the general results obtained by semi-classical
analysis; for results in arithmetic quantum chaos, we refer to
\cite{M}.

\section{Wave group and geodesic flow}

The model quantum Hamiltonians we will discuss are Laplacians
$\Delta$ on  compact Riemannian manifolds $(M,g)$ (with or without
boundary).  The classical  phase space in this setting  is the
cotangent bundle $T^*M$ of $M$, equipped with its canonical
symplectic form $\sum_i dx_i \wedge d\xi_i$. The metric defines
the Hamiltonian $H(x,\xi) = |\xi|_g =  \sqrt{\sum_{ij = 1}^n
g^{ij}(x) \xi_i \xi_j}$ on $T^*M$, where  $g_{ij} =
g(\frac{\partial}{\partial x_i},\frac{\partial}{\partial x_j}) $,
$[g^{ij}]$ is the inverse matrix to $[g_{ij}]$. We denote the
volume density of $(M, g)$ by $dVol$ and the corresponding inner
product on $L^2(M)$ by $\langle f, g \rangle$.  The unit (co-)
ball bundle is denoted $B^*M = \{(x, \xi): |\xi| \leq 1\}.$

 The  Hamiltonian flow $\Phi^t$ of $H$
is the geodesic flow. By definition, $\Phi^t(x, \xi) = (x_t,
\xi_t)$, where $(x_t, \xi_t)$ is the terminal tangent vector at
time $t$ of the unit speed geodesic starting at $x$ in the
direction $\xi$. Here and below, we often identify $T^*M$ with the
tangent bundle $TM$ using the metric to simplify the geometric
description. The geodesic flow preserves the energy surfaces $\{H
= E\}$ which are the co-sphere bundles $S^*_E M$.
 Due to the homogeneity of $H$, the
flow on any energy surface $\{H = E\}$ is equivalent to that on
the co-sphere bundle $S^*M = \{ H=1\}.$ (This homogeneity could be
broken by adding
 a potential $V \in C^{\infty}(M)$ to form a
semi-classical Schr\"odinger operator $- \hbar^2 \Delta + V$,
whose underlying Hamiltonian flow is generated by $|\xi|_g^2 +
V(x)$.)

The quantization of the Hamiltonian $H$ is the square root
$\sqrt{\Delta}$ of the positive Laplacian,  $$\Delta = -
\frac{1}{\sqrt{g}}\sum_{i,j=1}^n \frac{\partial}{\partial
x_i}g^{ij} g \frac{\partial}{\partial x_j} $$  of $(M,g)$. Here,
$g = {\rm det} [g_{ij}].$ We choose to work with $\sqrt{\Delta}$
rather than $\Delta$ since the former generates the wave
$$U_t =
e^{i t \sqrt{\Delta}}, $$ which is the quantization of the
geodesic flow $\Phi^t$.

 By the last statement  we mean that $U_t$ is related to
$\Phi^t$ in several essentially equivalent ways: \begin{enumerate}
\item  singularities of waves, i.e. solutions $U_t \psi$ of the
wave equation, propagate along geodesics; \item  $U_t$ is a
Fourier integral operator (= quantum map) associated to the
canonical relation defined by the graph of $\Phi^t$ in $T^*M
\times T^*M$;

\item  Egorov's theorem holds.

\end{enumerate}

We only define the latter since it plays an important role in
studying eigenfunctions. As with any quantum theory, there is an
algebra of observables on the Hilbert space $L^2(M, dvol_g)$ which
quantizes $T^*M$. Here, $dvol_g$ is the volume form of the metric.
The algebra is that $\Psi^*(M)$ of pseudodifferential operators
$\pdo$'s  of all orders, though we often restrict to  the
subalgebra $\Psi^0$ of $\pdo$'s of order zero.  We  denote by
$\Psi^m(M)$ the subspace of pseudodifferential operators of order
$m$. The algebra is defined by constructing a quantization $Op$
from an algebra of symbols $a \in S^m(T^*M)$ of order $m$
(polyhomogeneous functions on $T^*M \backslash 0)$ to $\Psi^m$.
The map $Op$ is not unique. In the reverse direction is the symbol
map $\sigma_A: \Psi^m \to S^m(T^*M)$ which takes an operator
$Op(a)$ to the homogeneous term $a_m$ of order $m$ in $a$. For
background we refer to \cite{HoIII, DSj}.

Egorov's theorem for the wave group  concerns the conjugations
\begin{equation} \label{ALPHAT} \alpha_t (A) : = U_t A U_t^*, \;\;\; A \in
\Psi^m(M). \end{equation} Such a conjugation defines the quantum
evolution  of observables  in the Heisenberg picture, and since
the early days of quantum mechanics it was known to correspond to
the classical evolution
\begin{equation}\label{VT}  V_t (a) : = a \circ \Phi^t \end{equation}
of observables $a \in C^{\infty}(S^*M)$. Egorov's theorem is the
rigorous version of this correspondence: it says that $\alpha_t$
defines an order-preserving automorphism of $\Psi^*(M)$, i.e.
$\alpha_t(A) \in \Psi^m(M)$ if $A \in \Psi^m(M)$, and that
\begin{equation} \sigma_{U_t A U_t^*} (x, \xi)  =  \sigma_A
(\Phi^t(x, \xi)) : = V_t (\sigma_A), \;\;\; (x, \xi) \in T^*M
\backslash 0.
\end{equation}
This  formula is almost universally taken to be the definition of
quantization of a flow or map in the physics literature.

The key difficulty  in quantum chaos is that it involves a
comparison between long-time dynamical properties of $\Phi^t$
 and $U_t$ through the symbol map and similar classical limits.
 The classical dynamics defines the `principal symbol' behavior of
 $U_t$ and the `error' $U_t A U_t^* - Op(\sigma_A \circ \Phi^t)$
 typically grows exponentially in time.  This is just the first
 example of a ubiquitous `exponential barrier' in the subject.

\section{Eigenvalues and eigenfunctions of $\Delta$}

The eigenvalue problem on a compact Riemannian manifold
$$\Delta \phi_j = \lambda_j^2 \phi_j,\;\;\;\;\;\;\;\;\langle \phi_j, \phi_k
\rangle = \delta_{jk}$$ is dual under the Fourier transform to the
wave equation. Here, $\{\phi_j\}$ is a choice of orthonormal basis
of eigenfunctions, which is not unique if the eigenvalues have
multiplicities $> 1.$ The individual eigenfunctions are difficult
to study directly, and so one generally forms the spectral
projections kernel,
\begin{equation} E(\lambda, x, y) = \sum_{j: \lambda_j \leq
\lambda} \phi_j(x) \phi_j(y). \end{equation} Semi-classical
asymptotics is the study of the $\lambda \to \infty$ limit of the
spectral data $\{\phi_j, \lambda_j\}$  or of $E(\lambda, x, y)$.
The (Schwartz) kernel of the wave group can be represented in
terms of the spectral data by
$$U_t (x,y) = \sum_j e^{it \lambda_j} \phi_j(x) \phi_j(y),$$
or equivalently as the Fourier transform $\int_{\R} e^{i t
\lambda} dE(\lambda, x, y)$ of the spectral projections. Hence
spectral asymptotics is often studied through the large time
behavior of the wave group.

The link between spectral theory and geometry, and the source of
Egorov's theorem for the wave group, is the construction of a
parametrix (or WKB formula) for the wave kernel. For small times
$t$, the simplest is the Hadamard parametrix,
\begin{equation} \label{HD} U_t (x,y) \sim \int_0^{\infty} e^{ i
\theta (r^2(x,y) - t^2)} \sum_{k = 0}^{\infty} U_k (x,y)
\theta^{\frac{d-3}{2} - k} d\theta\;\;\;\;\;\;\;\;\;(t < {\rm
inj}(M,g)) \end{equation} where $r(x,y)$ is the distance between
points, $U_0(x,y) = \Theta^{-\half}(x,y)$ is the volume
1/2-density, $inj(M, g)$ is the injectivity radius, and the higher
Hadamard coefficients are obtained by solving transport equations
along geodesics. The parametrix is asymptotic to the wave kernel
in the sense of smoothness, i.e. the difference of the two sides
of (\ref{HD}) is smooth. The relation (\ref{HD}) may be iterated
using $U_{t m} = U_t^m$ to obtain a parametrix for long times.
This is obviously complicated and not necessarily the best long
time parametrix construction, but it illustrates again the
difficulty of a long time analysis.

\subsection{Weyl law and local Weyl law}

A fundamental and classical result in spectral asymptotics is
Weyl's law on counting eigenvalues:
\begin{equation}\label{WL} N(\lambda ) = \#\{j:\lambda _j\leq \lambda \}= \frac{|B_n|}{(2\pi)^n} Vol(M, g) \lambda ^n
+O(\lambda ^{n-1}). \end{equation} Here, $|B_n|$ is the Euclidean
volume of the unit ball and $Vol(M, g)$ is the volume of $M$ with
respect to the metric $g$. An equivalent formula which emphasizes
the correspondence between classical and quantum mechanics is:
\begin{equation} Tr E_{\lambda} = \frac{Vol(|\xi|_g
\leq \lambda)}{(2\pi)^n}, \end{equation} where $Vol$ is the
symplectic volume measure relative to the natural symplectic form
$\sum_{j=1}^n dx_j \wedge d\xi_j$ on $T^*M$. Thus, the dimension
of the space where $H = \sqrt{\Delta} $ is $\leq \lambda$ is
asymptotically the volume where its symbol $|\xi|_g \leq \lambda$.

The remainder term in Weyl's law is sharp on the standard sphere,
where all geodesics are periodic, but is  not sharp on $(M, g)$
for which the set of periodic geodesics has measure zero
(Duistermaat-Guillemin \cite{DG}, Ivrii). When the set of periodic
geodesics, has measure zero (as is the case for ergodic systems),
one has
\begin{equation} \label{DGI} N(\lambda ) = \#\{j:\lambda _j\leq \lambda \}= \frac{|B_n|}{(2\pi)^n} Vol(M, g) \lambda ^n
+o(\lambda ^{n-1}). \end{equation} For background, see \cite{HoIV}
ch. XXIX. The remainder is then of small order than the derivative
of the principal term, and one then has asymptotics in shorter
intervals:
\begin{equation} \label{DGIshort} N([\lambda, \lambda + 1]) = \#\{j:\lambda _j  \in  [\lambda, \lambda + 1] \}=
n \frac{|B_n|}{(2\pi)^n} Vol(M, g) \lambda ^{n-1} +o(\lambda
^{n-1}).
\end{equation}
Physicists tend to write $\lambda \sim h^{-1}$ and to average over
intervals of this width. Then mean spacing between the eigenvalues
in this interval is  $\sim C_n Vol(M, g)^{-1} \lambda^{-(n-1)}$,
where $C_n$ is a constant depending on the dimension.

An important generalization is the {\it local Weyl law} concerning
the traces $Tr A E(\lambda)$ where $A \in \Psi^m(M).$ It asserts
that
\begin{equation} \label{LWL} \sum_{\lambda _{j} \leq
\lambda }\langle A\varphi_j, \varphi_j \rangle = \frac{1}{(2
\pi)^n}  \int_{B^*M} \sigma_A dx d\xi ) \lambda^n +
O(\lambda^{n-1}).
\end{equation}

There is also a  pointwise local Weyl law:

\begin{equation} \label{PLWL} \sum_{\lambda _{j} \leq
\lambda } |\varphi_j(x)|^2 = \frac{1}{(2 \pi)^n} |B^n| \lambda^n +
R(\lambda, x),
\end{equation}
where $R(\lambda, x) = O(\lambda^{n-1})$ uniformly in $x$. Again,
when the periodic geodesics form a set of measure zero in $S^*M$,
one could average over the shorter interval $[\lambda, \lambda +
1].$  Combining the Weyl and local Weyl law, we find the surface
average of $\sigma_A$ is a limit of traces:
\begin{equation} \label{OMEGA} \begin{array}{lll} \omega(A)
&: =&\displaystyle{ \frac{1}{\mu (S^{*}M)}\int_{S^*M}\sa d\mu} \\
& & \\
 &=& \displaystyle{ \lim_{\lambda
\rightarrow\infty}\frac{1}{N(\lambda )}\sum_{\lambda _{j} \leq
\lambda }\langle A\varphi_j, \varphi_j \rangle}
\end{array} \end{equation}

 Here,  $\mu$ is the {\it
Liouville measure} on $S^*M$, i.e. the  surface measure $d\mu =
\frac{dx d\xi}{d H}$  induced by the Hamiltonian   $H = |\xi|_g$
and by the symplectic volume measure $dx d\xi$
 on $T^*M$.

\subsection{Problems on asymptotics eigenfunctions}

Eigenfunctions arise in quantum mechanics as stationary states,
i.e. states $\psi$ for which the probability measure  $| \psi(t,
x)|^2 dvol$ is constant where $\psi(t, x) = U_t \psi(x)$ is the
evolving state. This follows from the fact that
\begin{equation} \label{UNIT} U_t \phi_k = e^{ i t \lambda_k} \phi_k \end{equation}
and that $|e^{i t \lambda_k}| = 1$.
 They are the basic modes of
the quantum system. One would like to know the behavior as
$\lambda_j \to \infty$ (or $\hbar \to 0$ in the semi-classical
setting) of invariants such as:

\begin{enumerate}

\item Matrix elements $\langle A \phi_j, \phi_j \rangle$ of
observables in this state;

\item Transition elements $\langle A \phi_i, \phi_j \rangle$
between states;

\item Size properties as measured by $L^p$ norms
$||\phi_j||_{L^p}$;

\item Value distribution as measured by the distribution function
$Vol\{x \in M: |\phi_j(x)|^2 > t\}. $

\item Shape properties, e.g. distribution of zeros and critical
points of $\phi_j$.

\end{enumerate}

Let us introduce  some  problems which have motivated much of the
work in this area. \medskip

\begin{prob} Let  ${\mathcal Q}$ denote the set of `quantum limits', i.e.
weak* limit points of the sequence $\{\Phi_k\}$ of distributions
on the classical phase space $S^*M$, defined by
$$\int_X a d\Phi_k := \langle Op(a) \phi_k, \phi_k \rangle$$
where $a \in C^{\infty}(S^*M)$.\end{prob}

The set $\qcal$ is independent of the definition of $Op$. It
follows almost immediately from Egorov's theorem that $\qcal
\subset \mcal_I$, where $\mcal_I$ is the convex set of invariant
probability measures for the geodesic flow. Furthermore, they are
time-reversal invariant, i.e. invariant under $(x, \xi) \to (x, -
\xi)$ since the eigenfunctions are real-valued.

To see this, it is helpful to introduce the linear functionals on
$\Psi^0$
\begin{equation} \label{RHOJ} \rho_k(A) = \langle Op(a) \phi_k, \phi_k
\rangle. \end{equation} We observe that $\rho_k(I) = 1$, that
$\rho_k(A) \geq 0$ if $A = \geq 0$ and that
\begin{equation}\label{INVAR}  \rho_k(U_t A U_t^*) = \rho_k(A). \end{equation}
Indeed, if $A \geq 0$ then $A = B^*B$ for some $B \in \Psi^0$ and
we can move $B^*$ to the right side. Similarly (\ref{INVAR}) is
proved by moving $U_t$ to the right side and using (\ref{UNIT}).
These properties mean that $\rho_j$ is an {\it invariant state} on
the algebra $\Psi^0$. More precisely, one should take the closure
of $\Psi^0$ in the operator norm. An invariant state is the
analogue in quantum statistical mechanics of an invariant
probability measure.

The next important fact about the states $\rho_k$ is that any weak
limit of the sequence $\{\rho_k\}$ on $\Psi^0$ is a probability
measure on $C(S^*M)$, i.e. a positive linear functional on
$C(S^*M)$ rather than just a state on $\Psi^0$. This follows from
the fact that $\langle K \phi_j, \phi_j \rangle \to 0$ for any
compact operator $K$, and so any limit of $\langle A \phi_k,
\phi_k \rangle$ is equally a limit of  $\langle (A + K)  \phi_k,
\phi_k \rangle$. Hence any limit is bounded by $\inf_K ||A + K||$
(the infimum taken over compact operators), and for any $A \in
\Psi^0$, $||\sigma_A||_{L^{\infty}} = \inf_K ||A + K||$. Hence any
weak limit is bounded by a constant times
$||\sigma_A||_{L^{\infty}} $ and is therefore continuous on
$C(S^*M)$. It is a positive functional since each $\rho_j$ is and
hence any limit is a probability measure. By Egorov's theorem and
the invariance of the $\rho_k$, any limit of $\rho_k(A)$ is a
limit of $\rho_k(Op (\sigma_A \circ \Phi^t))$ and hence the limit
measure is invariant.

Problem I is thus to identify which invariant measures in
$\mcal_I$ show up as weak limits of the functionals $\rho_k$ or
equivalently the distributions $d\Phi_k$. The weak limits reflect
the concentration and oscillation properties of eigenfunctions.
Here are some possibilities:

\begin{enumerate}

\item Normalized Liouville measure. In fact, the functional
$\omega$ of (\ref{OMEGA}) is also a state on $\Psi^0$ for the
reason explained above. A subsequence $\{\varphijk\}$ of
eigenfunctions is considered diffuse if $\rho_{j_k} \to \omega$.

\item A periodic orbit measure $\mu_{\gamma}$ defined by
$\mu_{\gamma}(A) = \frac{1}{L_{\gamma}} \int_{\gamma} \sigma_A ds$
where $L_{\gamma}$ is the length of $\gamma$. A sequence of
eigenfunctions for which $\rho_{k_j} \to \mu_{\gamma}$ obviously
concentrates (or strongly `scars') on the closed geodesic.

\item A finite sum of periodic orbit measures.

\item A delta-function along an invariant Lagrangian manifold
$\Lambda \subset S^*M$. The associated eigenfunctions are viewed
as {\it localizing} along $\Lambda$.

\item A more general measure which is singular with respect to
$d\mu$.

\end{enumerate}

All of these possibilities can and do happen in different
examples. If $d\Phi_{k_j} \to \omega$ then in particular, we have
$$\frac{1}{Vol(M)} \int_E |\phi_{k_j}(x)|^2 dVol \to
\frac{Vol(E)}{Vol(M)} $$ for any measurable set $E$ whose boundary
has measure zero. In the interpretation of $|\phi_{k_j}(x)|^2
dVol$ as the probability density of finding a particle of energy
$\lambda_k^2$ at $x$, this says that the sequence of probabilities
tends to uniform measure.

However,  $d\Phi_{k_j} \to \omega$ is much stronger since it says
that the eigenfunctions become diffuse on the energy surface
$S^*M$ and not just on the configuration space $M$. As an example,
consider the flat torus $\R^n/\Z^n$. An orthonormal basis of
eigenfunctions is furnished by the standard exponentials $e^{2 \pi
i \langle k, x \rangle}$ with $k \in \Z^n$. Obviously, $|e^{2 \pi
i \langle k, x \rangle}|^2 = 1$, so the eigenfunctions are already
diffuse in configuration space. On the other hand, they are far
from diffuse in phase space, and localize  on invariant Lagrange
tori in $S^*M$. Indeed, by definition of pseudodifferential
operator, $A e^{2 \pi i \langle k, x \rangle} = a(x, k) e^{2 \pi i
\langle k, x \rangle}$ where $a(x, k)$ is the complete symbol.
Thus, $$\langle A e^{2 \pi i \langle k, x \rangle}, e^{2 \pi i
\langle k, x \rangle} \rangle = \int_{\R^n/\Z^n} a(x, k) dx \sim
\int_{\R^n/\Z^n} \sigma_A(x, \frac{k}{|k|}) dx .
$$
A subsequence  $e^{2 \pi i \langle k_j, x \rangle}$ of
eigenfunctions has a weak limit if and only if $\frac{k_j}{|k_j|}$
tends to a limit vector $\xi_0$ in the unit sphere in $\R^n$. In
this case, the associated weak* limit is $\int_{\R^n/\Z^n}
\sigma_A(x, \xi_0) dx $, i.e. the delta-function on the invariant
torus $T_{\xi_0} \subset S^*M$ defined by the constant momentum
condition $\xi = \xi_0.$ The eigenfunctions are said to localize
under this invariant torus for $\Phi^t$.

The flat torus is a model of a completely integrable system, on
both the classical and quantum levels. Another example is that of
the standard round sphere $S^n$. In this case, the author and D.
Jakobson showed that absolutely any invariant measure $\nu \in
\mcal_I$ can arise as a weak limit of a sequence of
eigenfunctions. This reflects the huge degeneracy (multiplicities)
of the eigenvalues.

On the other hand, if the geodesic flow is ergodic one would
expect the eigenfunctions to be diffuse in phase space. In the
next section, we will discuss the rigorous results on this
problem.

Off-diagonal matrix elements
\begin{equation} \rho_{jk} (A) = \langle A \phi_i, \phi_j \rangle \end{equation} are also important
as  transition amplitudes between states. They  no longer define
states since $\rho_{jk}(I) = 0$, are no longer positive, and are
no longer invariant. Indeed, $\rho_{jk}(U_t A U_t^*) = e^{i t
(\lambda_j - \lambda_k)} \rho_{jk}(A),  $ so they are eigenvectors
of the automorphism $\alpha_t$ of (\ref{ALPHAT}). A sequence of
such matrix elements cannot have a weak limit unless the spectral
gap $\lambda_j - \lambda_k$ tends to a limit $\tau \in \R$. In
this case, by the same discussion as above, any weak limit of the
functionals $\rho_{jk}$ will be an eigenmeasure of the geodesic
flow which transforms by $e^{i \tau t}$ under the action of
$\Phi^t$. Examples of such eigenmeasures are orbital Fourier
coefficients $\frac{1}{L_{\gamma}} \int_0^{L_{\gamma}} e^{- i \tau
t} \sigma_A(\Phi^t(x, \xi)) dt$ along a periodic orbit. Here $\tau
\in \frac{2 \pi}{L_{\gamma}} \Z.$ We denote by $\qcal_{\tau}$ such
eigenmeasures of the geodesic flow. Problem $1$ has the following
extension to off-diagonal elements: \medskip

 \begin{prob} Determine the set
${\mathcal Q}_{\tau} $ of `quantum limits', i.e.  weak* limit
points of the sequence $\{\Phi_{kj}\}$ of distributions on the
classical phase space $S^*M$, defined by
$$\int_X a d\Phi_{kj} := \langle Op(a) \phi_k, \phi_j \rangle$$
where $\lambda_j - \lambda_k = \tau + o(1)$ and where $a \in
C^{\infty}(S^*M)$, or equivalently of the functionals
$\rho_{jk}$.\end{prob}

As will be discussed in \S \ref{QWMS}, the asymptotics of
off-diagonal elements depends on the weak mixing properties of the
geodesic flow and not just its ergodicity.

Matrix elements of eigenfunctions are quadratic forms. More
`nonlinear' problems involve the $L^p$ norms or the distribution
functions of eigenfunctions. Estimates of  the $L^{\infty}$ norms
can be obtained from  the local Weyl law (\ref{LWL}). Since the
jump in the the left hand side at $\lambda$ is $\sum_{j: \lambda_j
= \lambda} |\varphi_j(x)|^2$ and the jump in the right hand side
is the jump of $R(\lambda, x)$, this implies
\begin{equation} \label{SUP} \sum_{j: \lambda _{j} =
\lambda } |\varphi_j(x)|^2 = O(\lambda^{n-1}) \implies \;
||\varphi_j||_{L^{\infty} } = O(\lambda^{\frac{n-1}{2}}).
\end{equation}
For general $L^p$-norms, the following  bounds hold on any compact
Riemannian manifold \cite{Sog}:
\begin{equation}
\frac{\|\phi_j\|_p}{\|\phi\|_2}=O(\lambda^{\delta(p)}), \quad 2\le
p\le \infty.
\end{equation}
where
\begin{equation} \delta(p)=
\begin{cases}
n(\tfrac12-\tfrac1p)-\tfrac12, \quad \tfrac{2(n+1)}{n-1}\le p\le
\infty
\\
\tfrac{n-1}2(\tfrac12-\tfrac1p),\quad 2\le p\le
\tfrac{2(n+1)}{n-1}.
\end{cases}
\end{equation}

These estimates are sharp on the unit sphere $S^{n} \subset
\R^{n+1}$. The extremal eigenfunctions are the zonal spherical
harmonics, which are the $L^2$-normalized  spectral projection
kernels $\frac{\Pi_N(x, x_0)}{||\Pi_N(\cdot, x_0)||}$ centered at
any $x_0$. However, they are not sharp for generic $(M, g)$, and
it is natural to ask how `chaotic dynamics' might influence $L^p$
norms.

\begin{prob} Improve the estimates  $\frac{\|\phi_j\|_p}{\|\phi\|_2}=O(\lambda^{\delta(p)})$ for $(M, g)$
which ergodic or mixing geodesic flow. \end{prob}

In \cite{SogZ} it is proved that if a sequence of eigenfunctions
attains the bounds in (\ref{SUP}), then there must exist a point
$x_0$ so that a positive measure of geodesics starting at $x_0$ in
$S^*_{x_0} M$ return to $x_0$ at a fixed time $T$. In the real
analytic case, all return so $x_0$ is a perfect recurrent point.
In dimension $2$, such a perfect recurrent point cannot occur if
the geodesic flow is ergodic; hence $||\varphi_j||_{L^{\infty} } =
o(\lambda^{\frac{n-1}{2}})$ on any real analytic surface with
ergodic geodesic flow. This shows that none of the $L^p$ estimates
above the critical index are sharp for real analytic surfaces with
ergodic geodesic flow, and the problem is the extent to which they
can be improved.

The random wave model (see \S \ref{RWONB}) predicts that
eigenfunctions of Riemannian manifolds with chaotic geodesic flow
should have the bounds $||\phi_{\lambda}||_{L^p} = O(1)$ for $p <
\infty$ and that $||\phi_{\lambda}||_{L^{\infty}} < \sqrt{\log
\lambda}.$ But there are no rigorous estimates at this time  close
to such predictions. The best general estimate to date  on
negatively curved compact manifolds (which are models of chaotic
geodesic flow) is just the logarithmic improvement
$||\varphi_j||_{L^{\infty} } = O(\frac{\lambda^{n-1}}{\log
\lambda})$ on the standard remainder term in the local Weyl law.
This was known for compact  hyperbolic manifolds from the Selberg
trace formula, and similar  estimates hold manifolds without
conjugate points \cite{Ber}. The exponential growth of the
geodesic flow again causes a barrier in improving the estimate
beyond the logarithm. In the analogous setting of quantum `cat
maps', which are models of chaotic classical dynamics,  there
exist arbitrarily large eigenvalues with multiplicities of the
order $O(\frac{\lambda^{n-1}}{\log \lambda})$;  the
$L^{\infty}$-norm of the $L^2$-normalized projection kernel onto
an eigenspace of this multiplicity is of order the square root of
the multiplicity (\cite{FND}). This raises doubt that the
logarithmic estimate can be improved by general dynamical
arguments. Further discussion of $L^{\infty}$-norms, as well as
zeros,  will be given at the end of  \S \ref{QuE} for ergodic
systems.

\section{\label{QuE} Quantum ergodicity}

In this section, we discuss results on the problems stated above
when the geodesic flow of  $(M, g)$ is assumed to be ergodic. Let
us recall that this means that Liouville measure is an ergodic
measure for $\Phi^t$. This is a spectral property of the operator
$V_t$  of (\ref{VT}) on $L^2(S^*M, d\mu)$, namely that $V_t$ has
$1$ as an eigenvalue of multiplicity one. That is, the only
invariant $L^2$ functions (with respect to Liouville measure) are
the constant functions. This implies that the only invariant sets
have Liouville measure $0$ or $1$ and (Birkhoff's ergodic theorem)
that time averages of functions are constant almost everywhere
(equal to the space average).

 In
this case, there is a general result which originated in the work
of A. I. Schnirelman \cite{Sh.1, Sh.2}:

\begin{theo} \label{QE} \cite{Sh.1, Sh.2, Z0, Z.1, Z.3, CV, Su, GL, ZZw}  Let $(M,g)$ be a compact
Riemannian manifold (possibly with boundary), and let
$\{\lambda_j, \phi_j\}$ be the spectral data of its Laplacian
$\Delta.$ Then the geodesic flow
 $G^t$ is ergodic  on $(S^*M,d\mu)$ if and only if, for every
$A \in \Psi^o(M)$,  we have:
\medskip

\begin{enumerate}

 \item $\lim_{\lambda \rightarrow \infty}\frac{1}{N(\lambda)}
 \sum_{\lambda_j \leq \lambda}
|(A\phi_j,\phi_j)-\omega(A)|^2=0.$
\medskip

 \item $(\forall \epsilon)(\exists \delta)
\limsup_{\lambda \rightarrow \infty} \frac{1} {N(\lambda)}
\sum_{{j \not= k: \lambda_j, \lambda_k \leq \lambda}\atop {
|\lambda_j - \lambda_k| < \delta}} |( A \phi_j, \phi_k )|^2 <
\epsilon $
\end{enumerate}

\end{theo}

This implies that there exists a subsequence $\{\phi_{j_k}\}$ of
eigenfunctions whose indices $j_k$ have counting density one for
which $\langle A \phi_{j_k}, \phi_{j_k}\rangle \to \omega(A)$. We
will call the eigenfunctions in such a sequence `ergodic
eigenfunctions'. One can sharpen the results results by averaging
over eigenvalues in the   shorter interval $[\lambda, \lambda +
1]$  rather than in $[0, \lambda]$.

There is also an  ergodicity result for boundary values of
eigenfunctions on domains with boundary and with Dirichlet,
Neumann or Robin boundary conditions \cite{GL, HZ, Bu}. This
corresponds to the fact that the billiard map on $B^* \partial M$
is ergodic.

  The first
statement (1) is essentially a convexity result. It remains true
if one replaces the square by any convex function $\phi$ on the
spectrum of $A$,
\begin{equation} \label{CONVEX} \frac{1}{N(E)} \sum_{\lambda_j
\leq E} \phi (\langle A \phi_k, \phi_k\rangle  - \omega(A)) \to 0.
\end{equation}

Before sketching a proof, we  point out a somewhat heuristic
`picture proof' of the theorem.  Namely, ergodicity of the
geodesic flow is equivalent to the statement that Liouville
measure is an extreme point of the compact convex set $\mcal_I$.
In fact, it further implies that $\omega$ is an extreme point of
the compact convex set ${\mathcal E}_{\R}$ of invariant states for
$\alpha_t$ of (\ref{ALPHAT}); see \cite{Ru} for \S 6.3 for
background. But the local Weyl law says that $\omega$ is also the
limit of the convex combination $\frac{1}{N(E)} \sum_{\lambda_j
\leq E} \rho_j.$  An extreme point cannot be written as a convex
combination of other states unless all the states in the
combination are equal to it. In our case, $\omega$ is only a limit
of convex combinations so it need not (and does not) equal each
term. However, almost all terms in the sequence must tend to
$\omega$, and that is equivalent to (1).
\medskip

\noindent{\bf Sketch of Proof of (1)}~~~  As mentioned above, this
is a convexity result and  with no additional effort we can
consider more general  sums of the form
 We then have
\begin{equation} \sum_{\lambda_j \leq E} \phi (\langle A \phi_k, \phi_k\rangle
- \omega(A)) = \sum_{\lambda_j \leq E} \phi (\langle \langle A
\rangle_T - \omega(A) \phi_k, \phi_k\rangle ). \end{equation}  We
then apply the Peierls--Bogoliubov inequality
$$\sum_{j=1}^n \phi((B \phi_j, \phi_j)) \leq {\rm Tr\,} \phi (B)$$
with $B = \Pi_E [\langle A \rangle_T - \omega(A)]\Pi_E $ to get:
\begin{equation} \sum_{\lambda_j \leq E} \phi (\langle \langle A \rangle_T -
\omega(A) \phi_k, \phi_k\rangle  ) \leq {\rm Tr\,} \phi (\Pi_E
[\langle A \rangle_T - \omega(A)]\Pi_E ).\end{equation}

  Here, $\Pi_E$ is the
spectral projection for $\hat{H}$ corresponding to the interval
$[0, E].$  From the Berezin inequality [Si, (8.18)] we then have
(if $\phi(0) = 0$):
$$\frac{1}{N(E)} {\rm Tr\,} \phi (\Pi_E [\langle A \rangle_T - \omega(A)]\Pi_E ) \leq
\frac{1}{N(E)}  {\rm Tr\,} \Pi_E \phi ([\langle A \rangle_T -
\omega(A)]) \Pi_E \leqno(1.6.7)$$
$$ = \omega_E(\phi(\langle A
\rangle_T - \omega(A))).$$ As long as $\phi$ is smooth,
$\phi(\langle A \rangle_T - \omega(A))$ is a pseudodifferential
operator of order zero with principal symbol $\phi(\langle
\sigma_A \rangle_T - \omega(A)).$  By the assumption that
$\omega_E \rightarrow \omega$ we get
$$\lim_{E \rightarrow \infty}\frac{1}{N(E)} \sum_{\lambda_j \leq E} \phi
(\langle A \phi_k, \phi_k\rangle  - \omega(A)) \leq \int_{\{H =
1\}} \phi(\langle \sigma_A \rangle_T - \omega(A))
d\mu.\leqno(1.6.8)$$ As $T \rightarrow \infty$ the right side
approaches $\phi (0)$ by the dominated convergence theorem and by
Birkhoff's ergodic theorem.  Since the left hand side is
independent of $T$, this implies that
$$\lim_{E \rightarrow \infty}\frac{1}{N(E)} \sum_{\lambda_j \leq E} \phi
(\langle A \phi_k, \phi_k\rangle  - \omega(A)) = 0 $$ for any
smooth convex $\phi$ on ${\rm Spec}(A)$ with  $\phi (0) = 0.$ \qed
\bigskip

As mentioned above, the statement (1) is equivalent to saying that
there is a subsequence $\{\phi_{j_k}\}$ of counting density one
for which $\rho_{j_k} \to \omega$. The above proof does not and
cannot settle the question whether there exist exceptional sparse
subsequences  of eigenfunctions of density zero  tending to other
invariant measures. To see this, we observe that the proof is so
general that it applies to seemingly very different situations. In
place of the distributions  $\{\Phi_j\}$ we may consider the set
$\mu_{\gamma}$ of periodic orbit measures for a hyperbolic flow on
a compact manifold $X$. That is, $\mu_{\gamma} (f) =
\frac{1}{T_{\gamma}} \int_{\gamma} f $ for $f \in C(X),$ where
$\gamma$ is a closed orbit and $T_{\gamma}$ is its period.
According to the Bowen--Margulis equidistribution theorem for
closed orbits of hyperbolic flows, we have
$$\frac{1}{\Pi (T)} \sum_{\gamma: T_{\gamma} \leq T} \frac{1}{|{\rm det}(I -
P_{\gamma})|} \mu_{\gamma} \rightarrow \mu$$ where as above $\mu$
is the Liouville measure, where $P_{\gamma}$ is the linear
Poincar\'e map and where $\Pi (T) $ is the normalizing factor
which makes the left side a probability measure, i.e. defined by
the integral of $1$ against the sum.   An exact repetition of the
previous argument shows that up to a sparse subsequence of
$\gamma$'s, $ \mu_{\gamma} \rightarrow \mu$ individually. Yet
clearly, the whole sequence does not tend to $d\mu$: for instance
one could choose the sequence of iterates $\gamma^k$ of a fixed
closed orbit.
\bigskip

\subsection{Quantum ergodicity in terms of operator time and space
averages}

The first part of the result above may be reformulated as a
relation between operator time and space averages.

\noindent{\bf Definition}~~~{\it Let $A \in \Psi^0$ be an
observable and define its time average to be:
 $$\langle A \rangle := \lim_{T \rightarrow \infty} \frac{1}{2T} \int_{-T}^T
U_t^* A U_t dt$$ and its space average to be scalar operator
$$\omega (A) \cdot I$$ }
Here, the limit is taken in the weak operator topology (i.e. one
matrix element at a time).
 To see what is involved we
consider matrix elements with respect to the eigenfunctions.  We
have
$$( \frac{1}{2T} \int_{-T}^T U_t^* A U_t dt \phi_i, \phi_j) =
\frac{\sin  T(\lambda_i - \lambda_j)}{ T(\lambda_i - \lambda_j)}
(A \phi_i, \phi_j)\leqno(1.3.2)$$ from which it is clear that the
matrix element tends to zero as $T \rightarrow \infty$ unless
$\lambda_i = \lambda_j$.  However, there is no uniformity in the
rate at which it goes to zero since the spacing $\lambda_i -
\lambda_j$ could be uncontrollably small.

In these terms, Theorem \ref{QE} (1) says that:
\begin{equation} \langle A \rangle = \omega(A) I +
K,\;\;\;\;\;\;\mbox{where}\;\;\;\;\; \lim_{\lambda \rightarrow
\infty} \omega_{\lambda}(K^*K) \rightarrow 0, \end{equation} where
$\omega_{\lambda}(A) = Tr E(\lambda) A. $ Thus, the time average
equals the space average plus a term $K$ which is semi-classically
small in the sense that its Hilbert-Schmidt norm square
$||E_{\lambda} K||_{HS}^2$ in the span of the eigenfunctions of
eigenvalue $\leq \lambda$ is $o(N(\lambda)).$

This is not exactly equivalent to Theorem \ref{QE} (1) since it is
independent of the choice of orthonormal basis, while the previous
result depends on the choice of basis. However, when all
eigenvalues have multiplicity one, then the two are equivalent.
To see the equivalence, note that $\langle A \rangle$ commutes
with $\sqrt{\Delta}$ and hence is diagonal in the basis
$\{\phi_j\}$ of joint eigenfunctions of $\langle A \rangle$ and of
$U_t$. Hence $K$ is the diagonal matrix with entries $\langle A
\phi_k, \phi_k\rangle - \omega(A)$. The condition is therefore
equivalent to
$$\lim_{E \rightarrow \infty} \frac{1}{N(E)} \sum_{\lambda_j \leq E} |\langle A
\phi_k, \phi_k\rangle  - \omega(A)|^2 = 0.$$ Since all the terms
are positive, no cancellation is possible and this condition is
equivalent to the existence of a subset ${\mathcal S} \subset \N$
of density one such that ${\mathcal Q}_{{\mathcal S}} := \{ d
\Phi_k : k \in {\mathcal S}\}$ has only $\omega$ as a weak* limit
point. As above, one says that the sequence of eigenfunctions is
ergodic.

One could take this re-statement of Theorem \ref{QE} (1) as a
semi-classical definition of quantum ergodicity. Two natural
questions arise. First:

\begin{prob} Suppose the geodesic flow $\Phi^t$ of $(M, g)$ is ergodic on
$S^*M$. Is the operator $K$ in
$$\langle A \rangle = \omega(A) + K$$
 a compact operator? In this case,  $\sqrt{\Delta}$ is said to be QUE (quantum uniquely ergodic) If ergodicity
 is not sufficient for the QUE property, what extra conditions need to be
 added?  \end{prob}

Compact would imply that $\langle K \phi_k,  \phi_k \rangle \to
0$, hence $\langle A \phi_k, \phi_k \rangle \to \omega(A)$ along
the entire sequence.  Quite a lot of attention has been focussed
on this problem in the last decade. It is probable that ergodicity
is not by itself sufficient for the QUE property of general
quantum ergodic systems. For instance, it is believed  that there
exist modes of asymptotic bouncing ball type which concentrate on
the invariant Lagrangian cylinder (with boundary) formed by
bouncing ball orbits of the Bunimovich stadium (see e.g. \cite{KH}
for more on such `scarring'). Further, Faure-Nonnenmacher-de
Bi\`evre have shown that QUE does not hold for the hyperbolic
system defined by a quantum cat map on the torus \cite{FND}. Since
the methods applicable to eigenfunctions of  quantum maps and of
Laplacians have much in common,
 this negative result shows that there cannot exist a universal structural proof of
 QUE.

The principal positive result at this time is the proof by E.
Lindenstrauss \cite{Lin} of  the QUE property for the orthonormal
basis of Laplace-Hecke eigenfunctions eigenfunctions on arithmetic
hyperbolic surfaces.   It is generally believed that the spectrum
of  the Laplace eigenvalues is of multiplicity one for such
surfaces, so this should imply QUE completely for these surfaces.
Earlier  partial results on Hecke eigenfunctions  are due to
Rudnick-Sarnak \cite{R.S}, Wolpert \cite{W} and others. For more
on Hecke eigenfunctions, see \cite{M}.

So far we have not mentioned Theorem \ref{QE} (2). In the next
section we will describe a similar but more general result for
mixing systems and the relevance of (2) will become clear. An
interesting open problem is the extent to which (2) is actually
necessary for the equivalence to classical ergodicity.

\begin{prob} Converse QE:  What can be said of the classical limit of a
quantum ergodic system, i.e. a system for which $\langle A \rangle
= \omega(A) + K$ where $K$ is semiclassically in the sense above,
or compact?. Is it necessarily ergodic? \end{prob}

Very  little is known on this converse problem at present. It is
known that if there exists an open set in $S^*M$ filled by
periodic orbits, then the Laplacian cannot be quantum ergodic (see
\cite{MO} for recent results and references). But no proof exists
at this time that
 KAM systems, which have Cantor-like positive measure  invariant sets, are not
 quantum ergodic. It is known that there exist a positive proportion of approximate
eigenfunctions (quasi-modes) which localize on the invariant tori,
but  it has not been proved that a positive proportion of actual
eigenfunctions have this localization property.
\bigskip

\subsection{Further problems and  results on ergodic eigenfunctions}

Ergodicity is also known to have an impact on the distribution of
zeros. The  complex zeros in K\"ahler phase spaces of ergodic
eigenfunctions of quantum ergodic maps become uniformly
distributed with respect to the K\"ahler volume form  \cite{NV,
SZ}. An interesting problem is whether the real analogue is true:

\begin{prob} Ergodicity and equidistribution of nodal sets. Let
${\mathcal N}_{\phi_j} \subset M $ denote the nodal set (zero set)
of $\phi_j$, and equip it with its hypersurface volume form $d
{\mathcal H}^{n-1}$ induced by $g$. Let $(M, g)$ have ergodic
geodesic flow, and suppose that $\{\phi_j\}$ is an ergodic
sequence of eigenfunctions. Are the following asymptotics valid?
$$\int_{{\mathcal N}_{\phi_j}} f d {\mathcal H}^{n-1} \sim
\lambda_j \frac{1}{Vol(M, g)} \int_M f dVol. $$

\end{prob}

This is predicted by the random wave model of \S \ref{RWONB}.  An
equidistribution law for the complex zeros is known which gives
some evidence for the validity of this limit formula. Let $(M, g)$
be a compact real analytic Riemannian manifold and let
$\phi_j^{\C}$ be the holomorphic extension   of the real analytic
 eigenfunction $\phi_j$  to the
complexification $M_{\C}$ of $M$ (its Grauert tube).  Then if the
geodesic flow is ergodic and if $\phi_j$ is an ergodic sequence of
eigenfunctions, the  normalized current of integration
$\frac{1}{\lambda_j} Z_{\phi_j^{\C}}$  over the complex zero set
of $\phi_j^{\C}$ tends weakly to $\overline{\partial}
\partial |\xi_g|$. This current is invariant under the geodesic
flow and is singular along the zero section.

Finally, we mention some results on $L^{\infty}$ norms of
eigenfunctions on arithmetic hyperbolic manifolds of dimensions
$2$ and $3$. It is proved in \cite{IS} that the joint
eigenfunctions of $\Delta$ and the Hecke operators on arithmetic
hyperbolic surfaces  have the upper bound
$\|\phi_j\|_\infty=O_\epsilon(\lambda^{5/48+\epsilon}_j)$ for all
$j$ and $\epsilon>0$, and the lower bound $\|\phi_j\|_\infty\geq
c\sqrt{\log\log\lambda_j}$ for some constant $c>0$ and infinitely
many $j$. In \cite{R.S} it is proved that there exists an
arithmetic hyperbolic manifold and a subsequence $\phi_{j_k}$ of
eigenfunctions with $\|\phi_{j_k}\|_{L^\infty}\gg
\lambda^{1/4}_{j_k}$, contradicting  the  random wave model
predictions.

\section{\label{QWMS} Quantum weak mixing}

There are parallel results on quantizations of weak-mixing
geodesic flows which are the subject of this section. First we
recall the classical definition: the geodesic flow of $(M, g)$ is
weak mixing if the operator $V_t$ has purely continuous spectrum
on the orthogonal complement of the constant functions in
$L^2(S^*M, d\mu)$.  Hence like ergodicity it is a spectral
property of the geodesic flow.

We have:

\begin{theo} \label{QWM} ([Z.3,4])  The geodesic flow  $\Phi^t$ of $(M, g)$
is weak mixing if and only if the conditions (1)-(2) of Theorem
\ref{QE}  hold and additionally, for any $A \in \Psi^o(M)$,
$$(\forall \epsilon)(\exists \delta)
\limsup_{\lambda \rightarrow \infty} \frac{1} {N(\lambda)}
\sum_{{j\not= k: \lambda_j, \lambda_k \leq \lambda}\atop {
|\lambda_j - \lambda_k-\tau| < \delta}} |( A \phi_j, \phi_k )|^2 <
\epsilon \;\;\;\;\;\;\; (\forall \tau  \in \R )$$
\end{theo}

The restriction $j\not =k$ is of course redundant unless $\tau =
0$, in which case the statement coincides with quantum ergodicity.
This result follows from the general asymptotic formula, valid for
any compact Riemannian manifold $(M, g)$, that \begin{equation}
\label{QMF} \begin{array}{l}  \frac{1}{N(\lambda)}  \sum_{i \not=
j, \lambda_i, \lambda_j \leq \lambda} |\langle A \phi_i, \phi_j
\rangle|^2 \left|\frac{\sin T(\lambda_i -\lambda_j -\tau)}
{T(\lambda_i -\lambda_j -\tau)}\right|^2 \\ \\
 \sim ||\frac{1}{2T}
\int_{- T}^T e^{i t \tau} V_t(\sigma_A) ||_2^2 - |\frac{\sin T
\tau}{T \tau}|^2 \omega(A)^2. \end{array} \end{equation}  In the
case of weak-mixing geodesic flows, the right hand side $\to 0$ as
$T \to \infty$. As with diagonal sums, the sharper result is true
where  one averages  over the short intervals $[\lambda, \lambda +
1]$.

\subsection{Spectral measures and matrix elements}

Theorem \ref{QWM}  is based on expressing the spectral measures of
the geodesic flow in terms of matrix elements. The main limit
formula is:

\begin{equation} \label{SPECMEAS} \int^{\tau +\varepsilon }_{\tau-\varepsilon
} d\mu_{\sigma_A}:=\lim_{\lambda \rightarrow
\infty}\frac{1}{N(\lambda )}\sum_{i,j: \;\; \lambda _j\leq
\lambda,  \;\; |\lambda _i-\lambda _j-\tau|<\varepsilon
\\}\;
 |\langle A\varphi_i,
\varphi_j \rangle|^2\;\;,  \end{equation} where $d\mu_{\sigma_A}$
is the spectral measure for the geodesic flow corresponding to the
principal symbol of $A$,  $\sigma_A \in C^{\infty} (S^*M, d\mu)$.
Recall that the spectral measure of $V_t$ corresponding to $f\in
L^2$ is the measure $d\mu_f$ defined by
$$\langle V_tf,f \rangle_{L^2(S^*M)}  = \int_{\R} e^{\oit\tau} d\mu_f(\tau)\;.$$

The limit formula (\ref{SPECMEAS})  is equivalent to the dual
formula (under the Fourier transform)
\begin{equation} \label{QM2}\lim_{\lambda \to \infty} \frac{1}{N(\lambda)}   \sum_{i, j: \lambda_j \leq \lambda}
e^{i t (\lambda_i - \lambda_j)} |\langle A \phi_i, \phi_j
\rangle|^2 = \langle V_t \sigma_A, \sigma_A \rangle_{L^2(S^*M)}.
\end{equation}
The proof of (\ref{QM2}) is to consider, for $A\in \Psi^\circ$,
 the operator $A^*_tA\in \Psi^\circ$ with $A_t =
U^*_tAU_t$. By the local Weyl law,
$$\lim_{\lambda \rightarrow \infty}\frac{1}{N(\lambda )}
\Tr E(\lambda) A^*_tA = \langle V_t \sigma_A,\sigma_A
\rangle_{L^2(S^*M)}\;.$$ The right side of (\ref{SPECMEAS})
defines a measure $dm_A$ on $\R$ and (\ref{QM2}) says
$$\int_\R e^{it \tau}dm_A(\tau) = \langle V_t \sigma_A,\sigma_A
\rangle_{L^2(S^*M)}\;\;=\int_\R e^{it
\tau}d\mu_{\sigma_A}(\tau).$$

Since weak mixing systems are ergodic, it is not necessary to
average in both indices along an ergodic subsequence:

\begin{equation} \label{SPECMEASII} \lim_{\lambda_j \to \infty} \langle A_t^* A \phi_j, \phi_j \rangle =
 \sum_{j}
e^{i t (\lambda_i - \lambda_j)} |\langle A \phi_i, \phi_j
\rangle|^2 = \langle V_t \sigma_A, \sigma_A \rangle_{L^2(S^*M)}.
\end{equation}
Dually, one has

\begin{equation} \label{SPECMEASI} \lim_{\lambda_j \to \infty}
\sum_{i\; : \;  |\lambda _i-\lambda _j-\tau|<\varepsilon
\\}\;
 |\langle A\varphi_i,
\varphi_j \rangle|^2\;\; = \int^{\tau +\varepsilon
}_{\tau-\varepsilon } d\mu_{\sigma_A}.  \end{equation} For QUE
systems, these limit formulae are valid for the full sequence of
eigenfunctions.

%Equivalently, quantum  weak-mixing involves the Fourier
%coefficients
%\begin{equation} \hat A_T(\tau) =\frac{1}{2T} \int^T_{-T} e^{-it
%\tau}U^*_tAU_tdt
%\end{equation}  and the weak-limit
%$$\hat A(\tau) = \mbox{w-}\lim \frac{1}{2T}\int^T_{-T} e^{-it
%\tau}U^*_tAU_tdt\;.\leqno(1.7.2)$$ Just as $\bar A = \hat A(0)$ is
%the diagonal part of $A$ relative to $\{\varphi_j\}$, i.e.\ $\hat
%A(0) = [\langle A\varphi_i, \varphi_j\rangle\delta _{\lambda
%_{i},\lambda _{j}}]$, so is $\hat A(\tau) = [\langle A\varphi_i,
%\varphi_j\rangle\delta _{(\lambda _{i}-\lambda _{j}),\tau}]$.

\section{Rate of quantum ergodicity and mixing}

A quantitative refinement of quantum ergodicity is to  ask at what
rate the sums in Theorem \ref{QE}(1) tend to zero, i.e. to
establish a rate of quantum ergodicity. More generally, we
consider `variances' of matrix elements.  For diagonal matrix
elements, we define:
\begin{equation} \label{diag} V_A(\lambda) : =
\frac{1}{N(\lambda)} \sum_{j:  \lambda_j \leq \lambda} |\langle A
\phi_j, \phi_j) - \omega(A)|^2.
\end{equation}
In the off-diagonal case one may view $|\langle A\varphi_i,
\varphi_j \rangle|^2$ as analogous to $|\langle A \phi_j, \phi_j)
- \omega(A)|^2$. However, the sums in (\ref{SPECMEAS}) are double
sums while those of (\ref{diag}) are single. One may also average
over the shorter intervals $[\lambda, \lambda + 1].$

\subsection{Quantum chaos conjectures}

First, consider off-diagonal matrix elements. One conjecture is
that it is not necessary to sum in $j$ in (\ref{SPECMEASI}): each
individual term has the asymptotics consistent with
(\ref{SPECMEASI}). This is implicitly conjectured by
Feingold-Peres  in \cite{FP} (11) in the form
\begin{equation} \label{FPCONJ} |\langle A \phi_i, \phi_j
\rangle|^2 \simeq \frac{C_A (\frac{E_i - E_j)}{\hbar})}{2 \pi
\rho(E)},
\end{equation} where $C_A(\tau) = \int_{- \infty}^{\infty} e^{- i
\tau t} \langle V_t \sigma_A, \sigma_A \rangle dt. $ In our
notation, $\lambda_j = \hbar^{-1} E_j$ and $\rho(E) dE \sim d
N(\lambda)$. There are  $\sim C \lambda^{n-1}$ eigenvalues
$\lambda_i$ in the interval $[\lambda_j - \tau - \epsilon,
\lambda_j - \tau + \epsilon],$ so (\ref{FPCONJ}) says that
individual terms have the asymptotics of (\ref{SPECMEASI}).

On the basis of the analogy between $|\langle A\varphi_i,
\varphi_j \rangle|^2$ and  $|\langle A \phi_j, \phi_j) -
\omega(A)|^2$, it is  conjectured in \cite{FP}  that
$$\ V_A(\lambda)  \sim \frac{ C_{A - \omega(A) I}(0) }{\lambda^{n-1} vol(\Omega)}.
$$
The idea is that $\phi_{\pm} = \frac{1}{\sqrt{2}} (\phi_i \pm
\phi_j)$ have the same  matrix element asymptotics as
eigenfunctions when $\lambda_i - \lambda_j$ is sufficiently small.
But then $2 \langle A \phi_+, \phi_- \rangle = \langle A \phi_i,
\phi_i \rangle - \langle A \phi_j, \phi_j \rangle$ when $A^* = A$.
Since we are taking a difference, we  may replace each matrix
element by $\langle A \phi_i, \phi_i \rangle $ by $\langle A
\phi_i, \phi_i \rangle - \omega(A)$ (and also for $\phi_j$). The
conjecture then assumes that $ \langle A \phi_i, \phi_i \rangle -
\omega(A)$ has the same order of magnitude as $ \langle A \phi_i,
\phi_i \rangle - \langle A \phi_j, \phi_j \rangle$. Dynamical
grounds for this conjecture are given in  \cite{EFKAMM}. The order
of magnitude is predicted by some natural random wave models, as
discussed below in \S \ref{RWONB}.

\subsection{Rigorous results}

At this time, the strongest variance result is an asymptotic
formula for the diagonal variance proved by Luo-Sarnak for special
Hecke eigenfunctions on the quotient $\H^2/SL(2, \Z)$ of the upper
half plane by the modular group \cite{LS, Sa}. Their result
pertains to holomorphic Hecke eigenforms, but the analogous
statement for smooth Maass-Hecke eigenfunctions is expected to
hold by similar methods, so we state the result as a
Theorem/Conjecture. Note that $\H^2/SL(2, \Z)$ is a non-compact
finite area surface whose Laplacian $\Delta$ has both a discrete
and a continuous spectrum. The discrete Hecke eigenfunctions are
joint eigenfunctions of $\Delta$ and the Hecke operators $T_p$
(see \cite{Sa} for background).

\begin{theoconj} \cite{LS} \label{LS} Let $\{\phi_k\}$ denote the orthonormal basis of
Hecke eigenfunctions for $\H^2/SL(2, \Z)$. Then there exists a
quadratic form $B(f) $ on $C_0^{\infty}(\H^2/SL(2, \Z))$ such that
$$ \frac{1}{N(\lambda)}
 \sum_{\lambda_j \leq \lambda}
|\int_X f |\phi_j|^{2} dvol - \frac{1}{Vol(X)} \int_X f dVol|^2 =
\frac{B(f, f) }{\lambda} + o(\frac{1}{\lambda}).
$$
\end{theoconj}

 When the multiplier $f =
\phi_{\lambda}$ is itself an eigenfunction, Luo-Sarnak have shown
that
$$B(\phi_{\lambda}, \phi_{\lambda}) = C_{\phi_{\lambda}}(0)
L(\frac{1}{2}, \phi_{\lambda})$$ where $L(\frac{1}{2},
\phi_{\lambda})$ is a certain $L$-function. Thus, the conjectured
classical variance is multiplied by an arithmetic factor depending
on the multiplier. A crucial fact in the proof is that the
quadratic form B is diagonalized by the $\phi_{\lambda}$.

The only  rigorous result to date which is valid on general
Riemannian manifolds with hyperbolic geodesic flow  is the
logarithmic decay \cite{Z6}

\begin{theo} For any $(M, g)$ with hyperbolic geodesic flow,
$$ \frac{1}{N(\lambda)}
 \sum_{\lambda_j \leq \lambda}
|(A\phi_j,\phi_j)-\omega(A)|^{2p} = \frac{1}{(\log \lambda)^p}. $$
\end{theo}
The logarithm as usual reflects the exponential blow up in time of
remainder estimates for traces involving the wave group. It is
rather doubtful that such a result is sharp. However, in the case
of two-dimensional quantum cat maps, with eigenspaces of
multiplicty $\lambda/\log \lambda$, there may exist orthonormal
bases with rather large rates of ergodicity.

\section{\label{RWONB} Random waves and orthonormal bases}

We have mentioned that the random wave model provides a kind of
guideline for what to conjecture about eigenfunctions of quantum
chaotic system.  In this final section, we briefly discuss random
wave models and what they predict.

By a random wave model one means a probability measure on a space
of functions. To deal with orthonormal bases rather than
individual functions, one puts a probability measure on a space of
orthonormal bases, i.e. on a unitary group. We denote expected
values relative to a given probability measure by $\E$.  We now
consider some specific Gaussian models and what they predict about
variances.

 As a model for quantum chaotic eigenfunctions in plane domains, M. V. Berry suggested using the  {\it Euclidean
random wave model at fixed energy} \cite{B}.  Let
$\ecal_{\lambda}$ denote the space of (tempered) eigenfunctions of
eigenvalue $\lambda^2$ of the Euclidean Laplacian $\Delta$ on
$\R^n$. It is spanned by exponentials $e^{i \langle k, x \rangle}$
with $k \in \R^n, |k| = \lambda$. The infinite dimensional space
$\ecal_{\lambda}$ is a unitary representation of the Euclidean
motion group and carries an invariant inner product. The inner
product defines an associated Gaussian measure whose covariance
kernel $C_{\lambda}(x, y) = \E f(x) \bar{f}(y)$  is the derivative
at $\lambda$ of the spectral function
\begin{equation} E(\lambda, x, y) = (2 \pi)^{-n} \int_{|\xi| \leq \lambda} e^{i \langle x - y, \xi
\rangle } d \xi,\;\;\;\; \xi \in \R^n. \end{equation} Thus,
\begin{equation} \label{de0} C_{\lambda}(x, y) = \frac{d}{d \lambda} E(\lambda, x, y) = (2 \pi)^{-n} \int_{|\xi| = \lambda} e^{i \langle x - y, \xi
\rangle } d S = (2 \pi)^{-n} \lambda^{n-1} \int_{|\xi| = 1} e^{i
\lambda \langle x - y, \xi \rangle } d S, \end{equation} where
$dS$ is the usual surface measure. With this definition,
$C_{\lambda}(x, x) \sim \lambda^{n-1}$. In order to make
$\E(f(x)^2) = 1$ consistent with normalized eigenfunctions, we
divide by $\lambda^{n-1}$ to define
$$ \hat{C}_{\lambda}(x, y) =  (2 \pi)^{-n}  \int_{|\xi| = 1} e^{i
\lambda \langle x - y, \xi \rangle } d S. $$ One could express the
integral as a Bessel function to rewrite this as
$\Gamma(\frac{n-1}{2})
 |\lambda |x - y||^{-\frac{n-2}{2}} J_{\frac{n-2}{2}}(\lambda |x -
 y|)$.

 Wick's formula in this
ensemble gives:
$$\E \phi(x)^2 \phi(y)^2 = \frac{1}{Vol(\Omega)^2} [1 + 2
C_{\lambda}(x, y)^2]. $$ Thus, in dimension $n$ we have:
$$\begin{array}{lll} \E[ \int \int V(x) V(y) \phi(x)^2 \phi(y)^2 dx dy - \bar{V}^2] &= & \frac{2}{Vol(\Omega)^2}
\int_{\Omega} \int_{\Omega} \hat{C}_{\lambda}(x, y)^2
V(x) V(y) dx dy \\ & & \\
& \sim & \frac{1}{ \lambda^{n-1} Vol(\Omega)^2} \int_{\Omega}
\int_{\Omega} \frac{ V(x) V(y)}{|x - y|^{n-1}}   \cos (|x - y|
\lambda)^2 dx dy.
\end{array}$$
In the last line, we used the stationary phase asymptotics
\begin{equation}(2 \pi)^{-n} \lambda^{n-1} \int_{|\xi| = 1} e^{i
\lambda \langle x - y, \xi \rangle } d S\sim C_n (\lambda | x -
y|)^{-\frac{n-1}{
 2}}
\cos (|x- y|\lambda). \end{equation}  Thus, the variances have
order $\lambda^{-(n-1)}$ in dimension $n$, consistent with the
conjectures in \cite{FP, EFKAMM}.

This model is often used to obtain predictions on eigenfunctions
of chaotic systems. By construction it  is tied to Euclidean
geometry  and  only pertains directly to individual eigenfunctions
of a fixed eigenvalue. It is based on the infinite dimensional
multiplicity of eigenfunctions of fixed eigenvalue of the
Euclidean Laplacian on $\R^n$. There also exist random wave models
on a curved Riemannian manifold $(M, g)$, which model individual
eigenfunctions and also random orthonormal bases \cite{Z.2, Z.5}.
Thus, one can compare the behavior of sums over eigenvalues of the
orthonormal basis of eigenfunctions of $\Delta$ with that of a
random orthonormal basis.  Instead of taking Gaussian random
combinations of Euclidean plane waves of a fixed eigenvalue, one
takes Gaussian random combinations $\sum_{j: \lambda_j \in
[\lambda, \lambda + 1]} c_j \phi_j$  of the eigenfunctions of $(M,
g)$ with eigenvalues in a short interval in the sense above.
Equivalently, one takes random combinations with $\sum_{j} |c_j|^2
= 1$. These random waves are globally adapted to $(M, g)$.
 The statistical results
 depend on the measure of
the set of periodic geodesics of $(M, g)$; thus, as discussed in
\cite{KHZ}, different random wave models make different
predictions about off-diagonal variances.

%Analogous random wave models in the quantum maps setting are in
%\cite{BBL, NV}.

Fix a compact Riemannian manifold $(M, g)$ and partition the
spectrum of $\sqrt{\Delta}$ into the intervals $I_k = [k, k + 1]$.
Let $\Pi_k = E(k+1) - E(k)$ be the kernel of the spectral
projections for $\sqrt{\Delta}$ corresponding to the interval
$I_k$.  Its kernel $\Pi_k(x, y)$ is the covariance kernel of
Gaussian random combinations $\sum_{j: \lambda_j \in I_k} c_j
\phi_j$ and is analogous to $C_{\lambda}(x, y)$ in the Euclidean
case;  it is of course not the derivative $d E(\lambda, x, y)$ but
the difference of the spectral projector over $I_k$.   We denote
by $N(k)$ the number of eigenvalues
  in $I_k$ and  put $\hcal_k = \mbox{ran} \Pi_k$ (the range of
 $\Pi_k$). We define a  {\it random} orthonormal basis of $\hcal_k$ by
changing the basis of eigenfunctions $\{\phi_j\}$ of $\Delta$ in
$\hcal_k$ by a random element of the unitary group $U(\hcal_k)$ of
the finite dimensional Hilbert space $\hcal_k$.  We then define a
random orthonormal basis of $L^2(M)$ by taking the product over
all the spectral intervals in our partition. More precisely, we
define the infinite dimensional unitary group
$$U(\infty) =\Pi^{\infty}_{k=1}  U(\hcal_k)$$
 of  sequences
 $(U_1, U_2,\dots)$, with $U_k \in U(\hcal_k)$.
 We equip $U(\infty)$ with the
 product
$$d\nu_{\infty}= \Pi^{\infty}_{k=1} d\nu_k$$
 of the
 unit mass Haar measures $d\nu_k$ on $U(\hcal_k)$:
We then define  a  {\it random} orthonormal basis of $L^2(M)$ to
be obtained by applying a random element  $U \in U(\infty)$ to the
orthonormal basis $\Phi = \{\phi_j\}$ of eigenfunctions of
$\sqrt{\Delta}$.

Assuming the set of periodic geodesics of $(M, g)$ has measure
zero,  the Weyl remainder results (\ref{DGI}) and strong Szeg\"o
limit asymptotics of \cite{GO, LRS} give two term asymptotics  for
the traces $\Pi_k A \Pi_k, (\Pi_k A \Pi_k)^2$ for any
pseudodifferential operator $A$.  Combining  the strong Szeg\"o
asymptotics with the arguments of \cite{Z.5},  random orthonormal
bases can be proved to satisfy the following variance asymptotics:

$$\begin{array}{ll} (i) &  \E (
 \sum_{j: \lambda_j \in I_k}
|(AU\phi_j, U\phi_j)-\omega(A)|^2 \sim  ( \omega(A^*A) -
\omega(A)^2 ); \\ & \\(ii) &
  \E ( \sum_{i\not= j:
\lambda_j, \lambda_i \in I_k } \left|\frac{sin T(\lambda_i -
\lambda_j - \tau)} {T(\lambda_i - \lambda_j - \tau)}\right|^2 |( A
U\phi_j, U\phi_i )|^2 \\ & \\
&  \sim \{ 2 \left|\frac{sin\tau T}{\tau T}\right|^2 +
\frac{1}{N(k)} \sum_{i \not= j} \left|\frac{sinT(\lambda_i
-\lambda_j -\tau)} {T(\lambda_i -\lambda_j -\tau)}\right|^2 \}
(\omega(A^*A) - \omega(A)^2)
\end{array}$$

%What the random wave model conjectures assert is that the
%orthonormal basis of eigenfunctions of a mixing system behaves
%like a Gaussian random orthonormal basis formed by taking random
%combinations of the same eigenfunctions.

% Putting   $T = \log \lambda$ (the  Ehrenfest time) gives:
%$$\begin{array}{ll} (ii)' &   \E (\frac{1}{N(\lambda)} \sum_{{i\not= j:
%\lambda_j, \lambda_i \leq \lambda}\atop { |\lambda_j - \lambda_i
%-\tau| < \frac{1}{\log \lambda}}} |( A U\phi_j, U\phi_i )|^2 \sim
% \lambda \{\frac{(\sin
%\tau \log \lambda)^2}{( \tau \log \lambda)^2} \omega(A)^2\\ &\\ &
%+ \frac{1}{N(\lambda)^2} \sum_{\{i,j:\lambda_i,\lambda_j \leq
%\lambda\}}\left|\frac{sin \log \lambda (\lambda_i - \lambda_j -
%\tau)} {\log \lambda (\lambda_i - \lambda_j - \tau)}\right|^2
%\omega(A^2)\} \end{array}$$ for all $\tau \in \R$.

%$\clubsuit$ Need to fix. The rhs must depend on $N(k)$.
%$\clubsuit$


\begin{thebibliography}{HHHH}

\bibitem[A]{A} O. Agam, B. Altshuler and  A. V.  Andreev,
Spectral statistics: from disordered to chaotic systems. Phys.
Rev. Lett. 75 (1995), no. 24, 4389--4392

\bibitem[BSS]{BSS} A. B\"acker, R.  Schubert and P. Stifter,
Rate of quantum ergodicity in Euclidean billiards.  Phys. Rev. E
(3) 57 (1998), no. 5, part A, 5425--5447. ( Erratum: "Rate of
quantum ergodicity in Euclidean billiards". Phys. Rev. E (3) 58
(1998), no. 4, 5192).

\bibitem[Bar]{Bar} A. Barnett, Asymptotic rate of quantum
ergodicity in chaotic Euclidean billiards (2005).

\bibitem[Ber]{Ber} P. B\'erard,
On the wave equation on a compact Riemannian manifold without
conjugate points. Math. Z. 155 (1977), no. 3, 249--276.

\bibitem[B]{B} M. V. Berry,
Regular and irregular semiclassical wavefunctions. J. Phys. A 10
(1977), no. 12, 2083--2091.

\bibitem[BBL]{BBL} E. Bogomolny, O. Bohigas and P. Leboeuf, Quantum chaotic dynamics and random polynomials. J. Statist. Phys. 85 (1996), no. 5-6, 639--679.

\bibitem[BGS]{BGS} O. Bohigas, M. J. Giannoni and C.  Schmit,
Characterization of chaotic quantum spectra and universality of level fluctuation laws. Phys. Rev. Lett. 52 (1984), no. 1, 1--4.

\bibitem[Bu]{Bu} N. Burq,
Quantum ergodicity of boundary values of eigenfunctions: A control
theory approach, to appear in Canadian Math. Bull.
(math.AP/0301349).


\bibitem[CV]{CV} Y.Colin de Verdi\`ere, Ergodicit\'e et fonctions propres du
Laplacien, Comm.Math.Phys. 102 (1985), 497-502.

\bibitem[DSj]{DSj} M. Dimassi and J.  Sj\"ostrand, {\it Spectral asymptotics in the semi-classical limit}.
 London Mathematical Society Lecture Note Series, 268. Cambridge University Press, Cambridge, 1999.

 \bibitem[DG]{DG} J. Duistermaat and V. Guillemin, The spectrum of positive
elliptic operators and periodic bicharacteristics, Invent. Math.
29 (1975), 39--79.



 \bibitem[EFKAMM]{EFKAMM}  B. Eckhardt, S. Fishman, J. Keating, O. Agam, J. Main and K. Miller,
  Approach to Ergodocity in Quantum Wave Functions, Phys. Rev. E 52, 5893-5903 (1995).


\bibitem[FND]{FND} F. Faure, S. Nonnenmacher and S. De Bi\`evre, Scarred eigenstates for quantum cat maps of minimal periods. Comm. Math. Phys. 239 (2003), no. 3, 449--492

\bibitem[FP]{FP} M. Feingold and A. Peres,  Distribution of matrix elements of chaotic systems. Phys. Rev. A (3) 34 (1986), no. 1, 591--595.

\bibitem[GL]{GL} P.G\'erard and E.Leichtnam, Ergodic properties of
eigenfunctions for the Dirichlet problem, Duke Math J. 71 (1993),
559-607.

\bibitem[GO]{GO}  V. Guillemin and K. Okikiolu, Subprincipal terms in Szeg\"o
estimates, Math. Res. Lett. 4 (1997), 173--179.


\bibitem[Gu]{Gu} M. C. Gutzwiller, {\it Chaos in classical and quantum mechanics}. Interdisciplinary Applied Mathematics, 1. Springer-Verlag, New York, 1990

\bibitem[H]{H} E. J. Heller,
Bound-state eigenfunctions of classically chaotic Hamiltonian
systems: scars of periodic orbits. Phys. Rev. Lett. 53 (1984), no.
16, 1515--1518.

\bibitem[IS]{IS} H. Iwaniec and P.  Sarnak, $L\sp \infty$ norms of eigenfunctions of arithmetic surfaces. Ann. of Math. (2) 141 (1995), no. 2, 301--320.

\bibitem[KHZ]{KHZ} L. Kaplan and E. J.  Heller,
Weak quantum ergodicity.  Phys. D 121 (1998), no. 1-2, 1--18.
Appendix by S. Zelditch.

\bibitem[KH]{KH} L. Kaplan and E.J.  Heller,
Measuring scars of periodic orbits. Phys. Rev. E (3) 59 (1999),
no. 6, 6609--6628.


\bibitem[HZ]{HZ} A. Hassell and S. Zelditch, Quantum ergodicity and boundary values of
eigenfunctions, Comm.Math.Phys.   Volume 248, Number 1 (2004)  119
- 168.

\bibitem[HoIII]{HoIII} L. H\"ormander, {\it The Analysis of
Linear Partial Differential Operators III}, Grundlehren {\bf 274
}, Springer-Verlag (1985).

\bibitem[HoIV]{HoIV} L. H\"ormander, {\it The Analysis of
Linear Partial Differential Operators IV}, Grundlehren {\bf 275},
Springer-Verlag (1985).


\bibitem[LRS]{LRS} A. Laptev, D.  Robert and Yu.  Safarov,
Remarks on the paper of V. Guillemin and K. Okikiolu:
"Subprincipal terms in Szegö estimates" [Math. Res. Lett. 4
(1997), no. 1, 173--179. Math. Res. Lett. 5 (1998), no. 1-2,
57--61.

\bibitem[Lin]{Lin} E. Lindenstrauss, Invariant measures and
arithmetic quantum ergodicity, Annals of Math. (to appear).

\bibitem[LS]{LS} W.Luo and P.Sarnak, Quantum ergodicity of eigenfunctions on
${\rm PSL}_2(\Z) \backslash {\bf H}^2$, IHES Publ. 81 (1995),
207-237.

\bibitem[L.S.2]{L.S.2} W.Luo and P.Sarnak, Quantum variance for
Hecke eigenforms, Annales Scient. de l'\'Ecole Norm. Sup. 37
(2004),  p. 769-799.

\bibitem[MO]{MO} J. Marklof and S. O'Keefe,  Weyl's law and quantum ergodicity for maps with divided phase space.
 With an appendix "Converse quantum ergodicity" by Steve Zelditch. Nonlinearity 18 (2005), no. 1,
 277--304.

 \bibitem[M]{M} J. Marklof, Arithmetic quantum chaos, this
 Encyclopedia.

\bibitem[NV]{NV} S. Nonnenmacher and A. Voros, Chaotic eigenfunctions in phase space. J. Statist. Phys. 92 (1998), no. 3-4, 431--518.


\bibitem[R.S]{R.S} Z.Rudnick and P.Sarnak, The behaviour of eigenstates of
arithmetic hyperbolic manifolds, Comm.Math.Phys. 161 (1994),
195-213.

\bibitem[Ru]{Ru} D. Ruelle, {\it Statistical mechanics: Rigorous results}. W. A. Benjamin, Inc., New York-Amsterdam 1969


\bibitem[Sa.1]{Sa.1} P.Sarnak, Arithmetic quantum chaos, in {\it The Schur
Lectures (Tel Aviv, 1992)}, {\it Israel Mathematical Conference
Proc. Vol. 8} (1995), 183-236.

\bibitem[Sa]{Sa} P.  Sarnak, Spectra of hyperbolic surfaces. Bull. Amer. Math. Soc. (N.S.) 40 (2003), no. 4,
441--478.

\bibitem[SZ]{SZ} B. Shiffman and S.  Zelditch,
 Distribution of zeros of random and quantum chaotic sections of positive line bundles. Comm. Math. Phys. 200 (1999), no. 3,
 661--683.


\bibitem[Sh.1]{Sh.1} A.I.Shnirelman, Ergodic properties of eigenfunctions,
Usp.Math.Nauk. 29/6 (1974), 181-182.

\bibitem[Sh.2]{Sh.2} A.I.Shnirelman, On the asymptotic properties of
eigenfunctions in the region of chaotic motion, addendum to
V.F.Lazutkin, {\it KAM theory and semiclassical approximations to
eigenfunctions}, Springer (1993).

\bibitem[Sog]{Sog} C. D. Sogge, Concerning the $L^p$ norm of spectral clusters for
second-order elliptic operators on compact manifolds, J. Funct.
Anal. 77 (1988), 123--138.

\bibitem[SogZ]{SogZ} C. Sogge and S.  Zelditch,
Riemannian manifolds with maximal eigenfunction growth.  Duke
Math. J. 114 (2002), no. 3, 387--437


\bibitem[Su]{Su} T.Sunada,  Quantum ergodicity. {\it Progress in inverse spectral geometry}, 175--196, Trends Math., Birkhäuser, Basel, 1997.

\bibitem[W]{W} S. A. Wolpert, The modulus of continuity for $\Gamma\sb 0(m)\backslash{\Bbb H}$
semi-classical limits. Comm. Math. Phys. 216 (2001), no. 2,
313--323.

\bibitem[Z0]{Z0} S. Zelditch,
Uniform distribution of eigenfunctions on compact hyperbolic
surfaces. Duke Math. J. 55 (1987), no. 4, 919--941


\bibitem[Z.1]{Z.1} S.Zelditch,  Quantum ergodicity of $C^*$ dynamical systems,
Comm.Math.Phys. 177 (1996), 507-528.

\bibitem[Z.2]{Z.2}------------, Quantum ergodicity on the sphere, Comm.\ Math.\
Phys., {\bf 146} (1992), 61-71.

\bibitem[Z.3]{Z.3}------------, Quantum transition amplitudes for
classically ergodic or completely integrable systems, J. Fun.\
Anal. {\bf 94} (1990), 415-436.

\bibitem[Z.4]{Z.4} -----------, Quantum Mixing, J.Fun.Anal.140 (1996), 68-86.

\bibitem[Z.5]{Z.5} -----------, A random matrix model for quantum mixing,
Int.Math. Res.Not. 3 (1996), 115-137.

\bibitem[Z6]{Z6}  -----------, On the rate of quantum ergodicity. I. Upper bounds. Comm. Math. Phys. 160 (1994), no. 1, 81--92


\bibitem[ZZw]{ZZw} S.Zelditch and M.Zworski,  Ergodicity of eigenfunctions
for ergodic billiards,  Comm.Math. Phys. 175 (1996), 673-682.


\end{thebibliography}
\end{document}